\documentclass[a4j,10pt]{article}
\usepackage{graphicx}
\usepackage{amsmath,bm}
\usepackage{color}

\title{On the Optimum Geometry and Training Strategy for   Chemical Classifiers that Recognize the Shape of a Sphere }

\author{ Jerzy Gorecki$^*$, Konrad Gizynski and Ludomir Zommer,\\
Institute of Physical Chemistry, Polish Academy of Sciences, \\
Kasprzaka 44/52, 01-224Warsaw, Poland, \\
$^*$ jgorecki@ichf.edu.pl}

\begin{document}
\maketitle

\begin{abstract}
In this paper, we continue the discussion on database classifiers constructed with networks of interacting chemical oscillators. In our previous papers \cite{ gizy16,zomm19} we demonstrated that a small, regular network of  oscillators  can predict if three random numbers in the range $[0,1]$ describe a point located inside a sphere inscribed within the unit cube $[0,1] \times [0,1] \times [0,1]$ with the accuracy exceeding $80 \%$. The parameters of the network were determined using evolutionary optimization. Here we apply the same technique to investigate if the classifier accuracy for this problem can be improved by selecting a specific geometry of interacting oscillators. We also address questions on the optimum size of the training database for evolutionary optimization and on the minimum size of the testing dataset for objective evaluation of classifier accuracy.

\end{abstract}

\section{Introduction}

"Information makes the world go round" - this sentence characterizes a substantial part of human activity at the beginning of the XXI-st century \cite{energy}. The tremendous progress in semiconductor technologies has been continuously stimulating the interest in information processing strategies and in computing media. When the Moore law was formulated \cite{moore} probably nobody expected that it would be held for over five decades. It has been anticipated that this law finally breaks down, and human civilization needs other computing media to keep the progress. To answer the needs, studies on alternative information processing technologies (chemical reactions, optical systems, or quantum phenomena) have been continued for over 30 years \cite{gram98,calu00,adam07,adam17}. The research on chemical computing is strongly motivated by expectations to understand information processing by living organisms, believed to be based on chemical reactions \cite{hake02}.

In this Chapter, we are concerned with information processing using a chemical medium. 
The approach presented below has been inspired by the properties of a medium in which the Belousov-Zhabotinsky (BZ) reaction proceeds.  
The BZ-reaction has probably been the most studied chemical reaction for which a complex evolution is manifested \cite{bel59,zhab64,tyso94,epst98}. The reaction is the oxidation of an organic substrate by bromine compounds in an acidic environment and the presence of a catalyst. The BZ-reaction became famous because oscillations can be easily observed, as the changes in concentrations of the catalyst in different oxidation forms are reflected by the medium color. If the ferroine is used as the catalyst, then the medium is red when the reduced catalyst ($Fe(phen)_3^{2+}$) is dominant. The medium becomes blue for a high concentration of the catalyst in the oxidized form ($Fe(phen)_3^{3+}$). The reaction includes an autocatalytic production of the reaction activator ($H Br O_2$). If the medium is spatially distributed and if the diffusion of the activator is allowed, then the region corresponding to a high concentration of the activator can trigger the reaction around, and a pulse of the activator propagating in space can appear. 

The interest in a BZ-reaction as a medium for chemical information processing comes from the fact that its properties are similar to those observed for the nerve system \cite{hake02}. Using a spatially distributed medium, one can form channels where the propagation of excitation pulses is observed. These pulses interact (annihilate) one with another and can change their frequency on non-excitable junctions between channels \cite{siel02}. There are many strategies in which the BZ-medium can be applied for computing. Information can be coded in concentrations of reagents, in the spatial structures, or in the spatio-temporal evolution \cite{muzi14,muzi16}. An excitable chemical medium allows for easy realization of logic gates \cite{toth95,stei96,moto99,siel01,adam05}. In such gates, the input and output states are coded in the presence or the absence of an excitation at a selected point of the computing medium within a specific time interval. However, unlike the semiconductor gates that can reliably operate for years \cite{feyn96}, the chemical logic is not that robust. In typical experimental conditions using a nonlinear reaction-diffusion medium \cite{toth95,stei96,adam02}, the time of stable operation is measured in hours; thus, the bottom-up design of chemical computers does not seem to be productive. The most effective algorithms are obtained if the chemical information processing medium works in parallel. For example, it happens if reactions at different points are coupled by the diffusion. Two classical algorithms of reaction-diffusion computing belong to such a class. One of them is the prairie-fire algorithm, which allows finding the shortest path in the labyrinth using wave propagation in an excitable medium \cite{stei95,agla97}. The other is the Kuhnert algorithm for image processing with the light inhibited variant of the oscillatory Belousov-Zhabotinsky reaction \cite{kuhn86, kuhn89, ramb97}. 

If the ruthenium complex (Ru(bpy)$_3$) is used as the catalyst, then the BZ reaction is photosensitive, and illumination with the blue light produces $Br^-$ ions that inhibit the reaction \cite{kuhn86,krug90,kada97}. After illumination of such an oscillatory medium, excitations are rapidly damped, and the system reaches a stable, steady state. On the other hand, the oscillatory behavior re-appears immediately after the illumination is switched off~\cite{gizy17b}. The existence of such external control is essential  
for information processing applications because it allows inputting information into the computing medium \cite{kuhn86, ramb97,yosh09}. 
For the analysis presented below, it is sufficient to assume that the controlling factor has an inhibiting effect.  The oscillations can be terminated when it is applied, and the oscillatory behavior is quickly restored when illumination is switched off. In this Chapter, following the analogy with the photosensitive BZ-reaction, we use the word illumination to describe the factor that controls oscillators. 

A number of reports on the information processing potential of networks composed of interacting chemical BZ-oscillators have appeared in the recent decade \cite{szym11,holl11,kuze19,vana20}. Our experiments have shown that networks of oscillators formed by touching droplets containing water solution of reagents of an oscillatory BZ-reaction 
can be stabilized by lipids dissolved in the surrounding oil phase. High uniformity of droplets that form the structure can be achieved if droplets are formed in a microfluidic device \cite{guzo16}. The touching droplets can communicate via the exchange of the activator. 
If phosphorolipids (asolectin) are used, then BZ droplets communicate mainly via the exchange of reaction activator ($H_2 Br O_3$ molecules), that can diffuse through the lipid bilayers \cite{szym11} and transmits excitation.

 The experiments indicate that  during a single oscillation cycle of a typical chemical oscillator we
can distinguish three phases: excited, refractive, and responsive \cite{epst98}.
Such distinction is important for the simplified model of interactions between
two oscillators coupled by the exchange of reaction activator. The excited
phase denotes the peak of activator concentration. An excited oscillator is
able to spread out activator molecules and to speed up their production in
the medium around. In the refractory phase, the concentration of inhibitor is
high, and in this phase, the oscillator does not respond to activator transported
from neighboring oscillators. In the responsive phase, the concentration of the
inhibitor decreases. An oscillator in this phase can get excited by interactions
with an oscillator in the excitation phase. These properties are reflected by the event-based-model that allows for a fast simulation of 
the evolution of large oscillator networks. The event-based-mode is illustrated in Fig. 1. 

The results presented below are based on numerical simulations of interacting chemical oscillators. The simplest mathematical model of BZ- reaction describes the reaction as an interplay between two reagents: the activator (HBrO2 molecules) and the inhibitor (the oxidized form of the catalyst). Two variable models like Oregonator \cite{fiel74} or Rovinsky-Zhabotinsky model  \cite{rovi84} give a pretty realistic description of simple oscillations, excitability, and the simplest spatio-temporal phenomena. However, the numerical complexity of models based on kinetic equations is still substantial, and they are too slow for large-scale simulations based on evolutionary optimization. Here, following \cite{gizy16} we use the event-based model schematically illustrated in Fig. 1 to simulate the time evolution of the medium. Following the previous papers \cite{gizy16,zomm19} we assumed that during the oscillation cycle, a droplet could be in one of there phases: the excitation phase lasting 1 second, the refractive phase, lasting 10 seconds or the responsive phase that is 19 seconds long. For an insolated oscillator, the excitation phase appears just after the responsive phase ends and the cycle repeats. Thus the period of the cycle is $30$ seconds. Oscillations with such period have been observed in experiments with BZ-medium \cite{gizy17b}. The separation of the oscillation cycle into refractory, responsive, and excitation phases allows introducing a simple model for interaction between droplets.  We assumed that if an oscillator is excited, then, 1 time unit later, all its neighbors in the responsive phase switch into the excitation phase. We also assume that immediately after illumination is switched on, the oscillator state changes into the refractory one. When the illumination is switched off, the excitation phase starts immediately.

\begin{figure}
\centerline{\includegraphics[width=10cm]{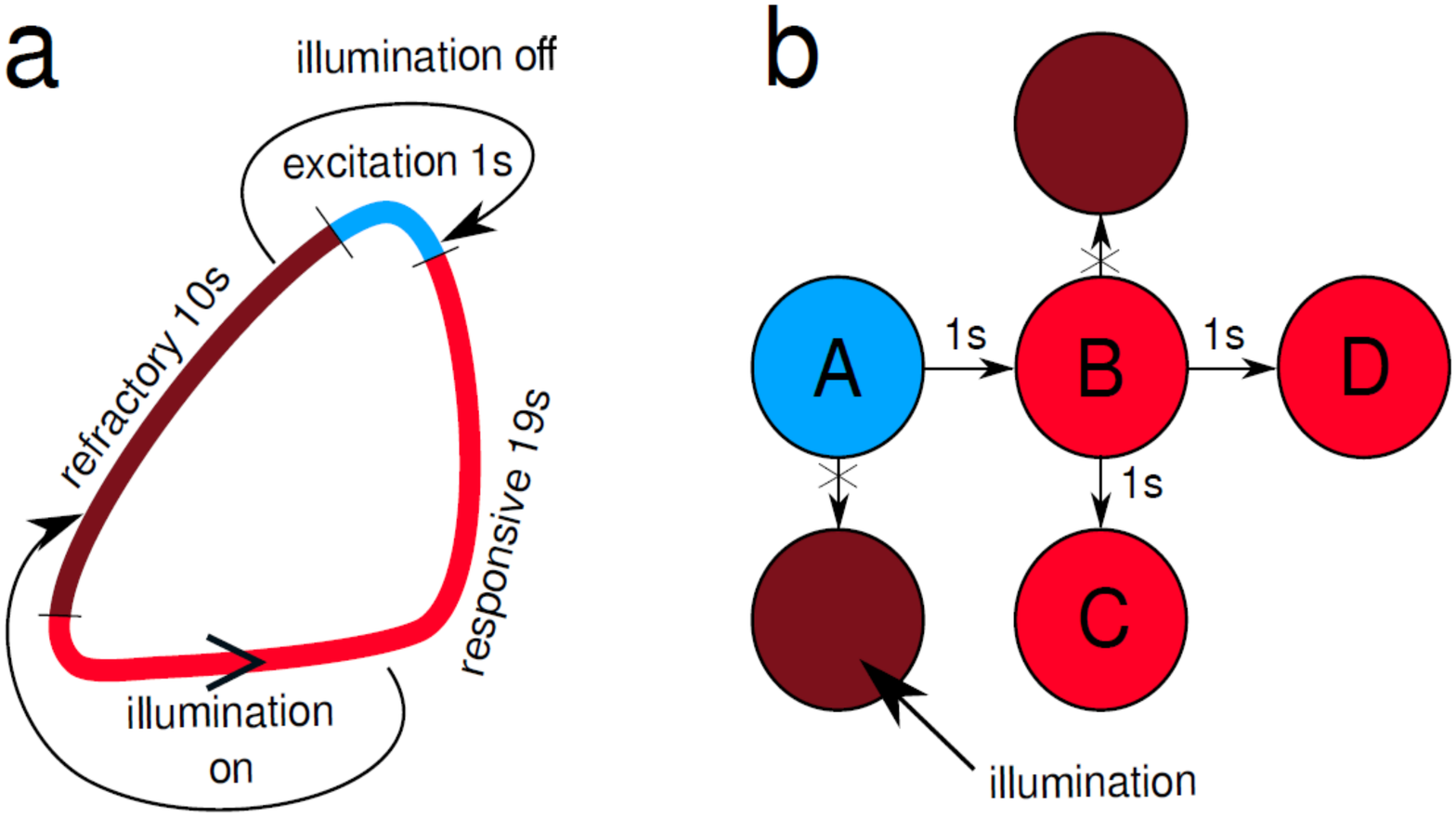} }
\caption{Graphical illustration of the event-based-model used  to simulate the time evolution of a single oscillator and the interactions between them in the network:
(a) The lengths of cycle phases (excitation, refractory and responsive) and the changes between phases when the illumination is switched on or off. (b) Interactions between the nearest neighbor oscillators: if an oscillator (A) is in the excitation phase, then,  $\Delta t=1$ s later, the excitation phase starts in all neighboring oscillators that are in the responsive phase.  Oscillators that are in the refractory state do nor change its state (reprinted form \cite{gizy16}).
}
\label{ra_fig1}
\end{figure}

In the following, we are concerned with the application of a  network of interacting chemical oscillators \cite{grue13,grue15} for database classification. In the considered classification problems, the database has the form of a set of records.
Each record is a $(n+1)-$tuple $(p_1, p_2,..., p_n, rt )$ where $p_i$ are predictors 
represented by real or integer numbers and $rt$ is an integer defining the record type. The classification program is supposed to output the correct record type if the predictor values are used as the input.  

It has been recently demonstrated that many information processing tasks can be performed by networks of interacting chemical oscillators using their specific
properties  \cite{gizy17a,gizy17c}. The idea of a dataset classifier based on a network of photosensitive chemical oscillators is illustrated in Fig.2.  The considered networks were formed by two types of oscillators (here denoted by circles): "normal" and "input" ones.
It is assumed that each oscillator in the network can be individually inhibited by an external factor. We can use this factor to introduce the input information and to control the evolution of the medium. The illuminations of input oscillators are related by an affine function with the predictor value of a given record. For normal oscillators, their illuminations are fixed. The normal oscillators are supposed to moderate interactions in the medium and to optimize them for a specific problem. In figures, the illumination time for normal oscillators is indicated with the density of a blue color.  It is also assumed that the output information about the record type can be extracted from the number of excitations (the number of maxima of a specific reagent concentration) observed at a selected set of oscillators within a fixed interval of time. In such an approach, information processing is a transient phenomenon. It does not matter if the system approaches a stationary state over a long time or not. 
The classification algorithm, obtained as the result of evolutionary optimization, consists of locations of input droplets and the output one and the set of illumination times applied to all normal droplets, regardless of the input value.  Our recent results suggest that reasonably accurate database classifiers can be constructed with a small network of coupled chemical oscillators \cite{gore20a,gore20b}.

\begin{figure}
\centerline{\includegraphics[width=10cm]{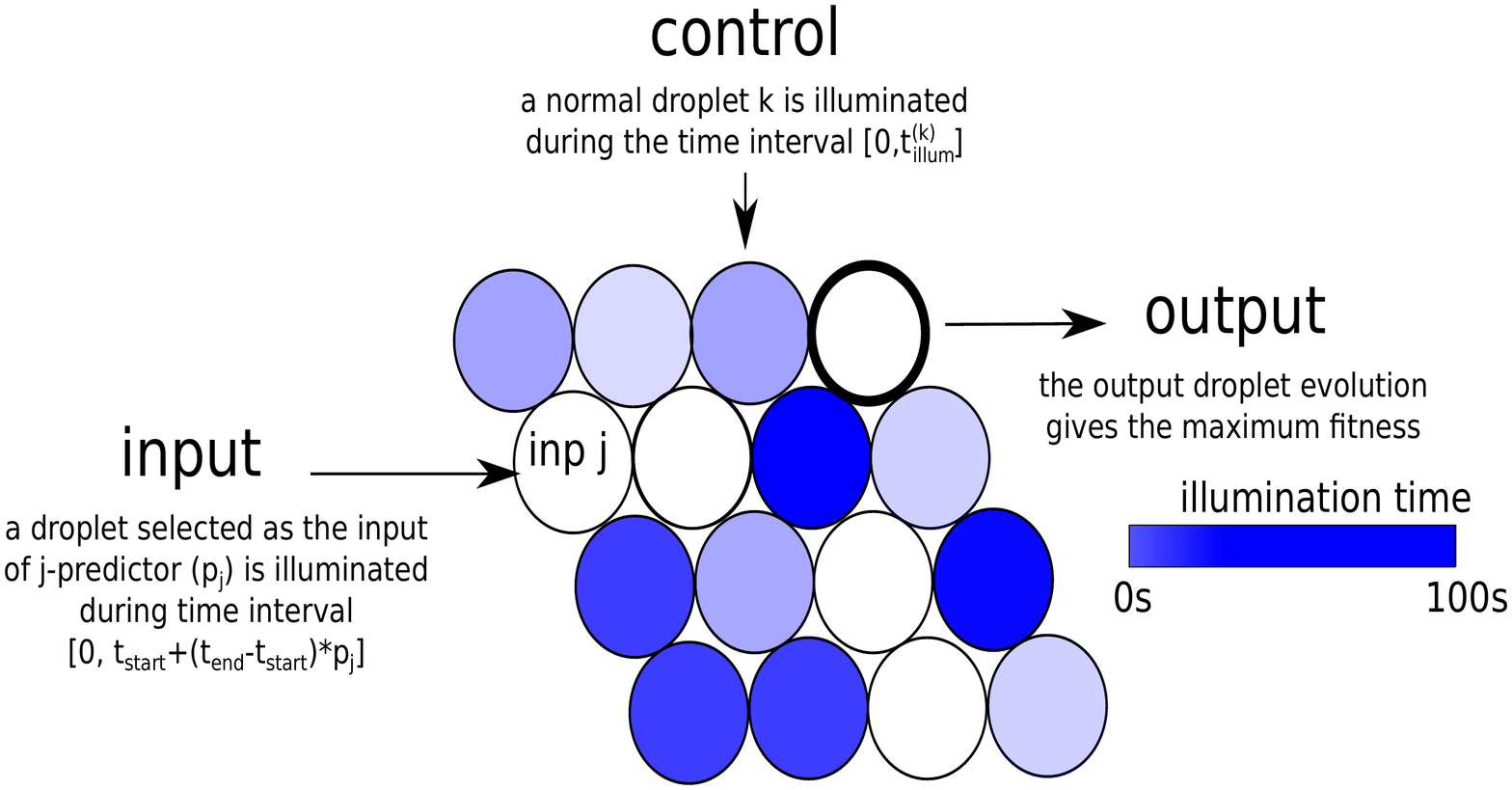} }
\caption{ 
The idea of a database classifier constructed with a network of chemical oscillators. We assume that touching oscillators interact. The classifier is made of oscillators of two types: "normal" and "input" ones. The classification “program”, obtained as the result of evolutionary optimization, consists of locations of input oscillators and the set of inhibition times applied to all normal oscillators, independent of the input value.  An input is provided by applying an inhibiting factor (illumination) to selected input oscillators, by the time related by an affine function with the predictor value. The number of excitations appearing at one oscillator, selected as the output one, is the classifier answer. Here and in the following figures, the output oscillator is marked by the thick border. The density of the blue color on normal oscillators increases with the time for which a given oscillator is inhibited.
}
\label{ra_fig2}
\end{figure}

The results presented below are the continuation of our previous papers \cite{gizy16,zomm19} in which we demonstrated that  networks of droplets containing reagents of Belousov-Zhabotinsky reaction can determine with reasonably high accuracy if three random numbers in the range $[0,1]$ describe a point located inside the Sphere In the Cube  $[0,1] \times [0,1] \times [0,1]$ (the SIC problem).  Here we consider a similar classification problem. The difference is in the sphere radius. In the  paper  \cite{gizy16} we considered the sphere inscribed in the unit cube with the radius $r=0.5$. The volume of such a sphere is $ 0.524$. In the following paper, \cite{} and here we consider a bit more complex problem of a centrally located sphere with a volume equal to $0.5$. The radius of such sphere is $r=( 3/(8 \pi) )^{1/3} \sim 0.492 $.  Now, the Shannon information of an unbiased set of points located inside and outside the sphere is 1 bit.  Datasets of this classification problem  are composed of records in the form of 4-tuples $(p_1, p_2, p_3, rt )$ where $p_i$ are x-, y- and z- coordinates of a point and the value of $rt$ is 1 if the point is inside the sphere ($(p_1-0.5)^2+( p_2-0.5)^2+( p_3-0.5)^2 < r^2$ ) and $rt = 0$ if the point i located outside the sphere. The SIC problem may seem academic but has some advantages. Unlike the database concerned with real problems, we can easily generate a database for the SIC with any number of elements. Therefore, we can investigate how the classifier accuracy depends on the number of records included in the process of its optimization. Moreover, using different databases, we can verify if a classifier optimized with one database retains its accuracy if another database of the same problem is used. 

In the reference  \cite{gizy16}, we presented results for the sphere inscribed in the cube  problem considering networks for which oscillators were located at nodes of a 2-dimensional regular network and interacted with the nearest neighbors. We considered different numbers of oscillators; from 4 ($ = 2 \times 2$) to 25 ($ = 5 \times 5$). The classifier accuracy was estimated by counting the number of correct answers for the datasets used during the classifier optimization. As expected, large classifiers performed better, and their accuracy was an increasing function of the classifier size. The highest measured accuracy was 87.5\% for 4x4 classifier trained on the dataset of 200 cases. For training datasets composed of 400 records the accuracy was an increasing function and it changed from $77 \%$ of correct answers for $ = 2 \times 2$ network to $81 \%$ for $ = 5 \times 5$ network. Thus, the increase in accuracy with the network size was quite small. In  the reference \cite{zomm19} we considered the SIC problem with the sphere volume equal to $0.5$ unit. Nine oscillators of the network were arranged in a hexagonal geometry with interactions between the nearest oscillators. The accuracy of optimized classifiers was close to $ 90 \%$. It was a substantial increase if compared with networks in a regular geometry.

Here we continue the study on chemical classifiers of Sphere In the Cube problem and answer questions that have not been addressed in the previous paper: First, there is a question if the same dataset should be used for classifier optimization and for testing its accuracy. Having in mind practical applications of chemical classifiers, we expect that after proper training, the classifier should correctly treat all cases, not only those that were included in the process of its training as it was done in \cite{gizy16}. Here we introduce independent databases of different sizes and estimate what number of records is necessary for objective estimation of classifier accuracy. We discuss the problem of optimization effectiveness by considering different population sizes in the evolutionary algorithm and different numbers of elements in the training datasets. We also consider networks of different geometries (cf. Fig. 7 ) in order to verify which network geometry can improve classifier accuracy. 

The Chapter is arranged into four Sections. The second Section contains basic information on the genetic algorithm used for classifier optimization. A reader who is not familiar with the subject is advised to read our previous paper \cite{gizy16} where the details of the methods are given. In the third Section we present results for oscillator-based classifiers of the Sphere in a Cube problem for different geometries of the network, different optimization parameters, and different sizes of datasets used for the optimization procedure. In the final Section we discuss the obtained results.

\section{The evolutionary optimization of chemical classifiers}

It has been demonstrated \cite{gizy17a,gizy17c,gore20a} that the top-down approach based on the evolutionary optimization can be successfully applied to design classifiers based on coupled chemical oscillators.  At the beginning, we have to fix the time interval $[0, t_{max}]$ within which the time evolution of the network evolution is observed, and the numbers of excitation on all oscillators are counted. All reported results were obtained for  $t_{max} = 100 s$. This is an important assumption because it says that information processing is a transient phenomenon. We assume that output information can be extracted by observing the system within the time interval $[0, t_{max}]$ and it does not matter if a network reaches a steady state after a long time.

Within the top-down approach, first, we specify the function that should be performed by the considered system. Next, we search for possible factors that can modify the system, increasing its information processing ability. Finally, we combine all these factors and apply them to achieve optimum performance. Here, (cf. Fig. 2)  these factors are the locations of the input and the normal oscillators and the illumination intervals for all intervals. We assume that all oscillators were inhibited at the time $t=0$ and the oscillator \#i remained inhibited within the time interval $[0, t_{illum}(i)]$. For normal oscillators, the illumination times $ t_{illum}(i)$ are fixed and the same for all processed records. If an oscillator \#i  is considered as the input one for the $j$-th predictor, and the predictor value in the record \#k  is $p_j \in[0,1]$ than, when processing the record \#k  this oscillator is inhibited (illuminated) within the time interval $[0,  t_{start} + (t_{end} - t_{start})*p_j]$ where the values of  $t_{start}  \le t_{max}$ and  $t_{end}  \le t_{max}$  are the same for all predictors. Therefore, when processing the record \#k,  $ t_{illum}(i) =t_{start} + (t_{end} - t_{start})*p_j  $.  In the following we assume that $t_{start}$ and $t_{end}$ are identical for all predictors. The symmetry of the considered problem can justify such an assumption.

Therefore,   a network classifier can be defined if the folowing parameters are defined: $Geom$ - the network geometry and interactions between droplets,  $t_{max}$, $Loc$ - location  
of the input droplets, $t_{end}$, $t_{start}$ and $ t_{illum}(i)$.
Of course, it would be naive to believe that a randomly selected network of oscillators performs a 
correct classification of any database we choose. The network parameters should be optimized according to the problem we are going to solve. To do it, we select a training database of the problem and perform a complex, multi-parameter optimization of the network using an evolutionary algorithm. In the beginning, we selected a training dataset of $K$ records and a population of $M$ classifiers with randomly initialized parameters.
Next, the fitness of each classifier was evaluated. 

In the previous papers \cite{gizy16,gizy17a,gizy17c,zomm19} we measured the fitness using the mutual information between the set of record types and the set of the number of excitation observed on oscillators.
The mutual information between two sets \cite{cove06} is the quantity that describes how much information about an element of one set can be gained if we know the element of the other. Let us assume that the training dataset is D and consider its classifier C. For each record $r$ ($r \in D$) characterized by the record type $rt$ we can detect the number of excitations $o_{rj}$ observed on the oscillator $j$ of the network. For each oscillator $j$ of the classifier C we calculate  the mutual information between the sets $R=\{ rt, r \in D \}$ and $O_j=\{ o_{rj}, r \in D \}$. The value $I(R,O_j)$ is:\\
$I(R,O_j) = (H(R) + H(O_j) - H(R, O_j)  $, \\
where $H(R)$ and $H(O_j)$ are the Shannon entropies \cite{shan48} of the set of output types in the training dataset and of the set of observed on the droplet $j$ of the network and $H(\mathcal{S}_i,\mathcal{P}_o)$ is the joint entropy of both these sets. We select the output oscillator as the one for which $I(R,O_j)$ has the maximum. The value $I(R,O) = max_j I(R,O_j)$ defines the fitness of the classifier C.

Alternatively, we can use the classifier accuracy as the measure of its fitness. Let us consider the oscillator \#j  and a particular number of its excitations $o$. We can introduce two sets:  $O^0_j=\{r; o_{rj}=o \land r=0, r \in D \}$ and 
$O^1_j=\{r; o_{rj}=o \land r=1, r \in D \}$. If the set $O^0_j$ has more elements than the set $O^1_j$ then it is more likely to that the record corresponding to a point outside the sphere is processed when  $o$ excitations at the oscillator $j$ are observed.  If the set $O^1_j$ has more elements than the set $O^0_j$ then the probability that the considered record describes a point inside the sphere is higher. Considering all possible numbers of excitation observed on the oscillator $j$ we can calculate the accuracy of classification $A_j$ if the oscillator $j$  is used as the output one. The maximum $ max_j A_j$ the classifier accuracy and the oscillator $j$ at which it is observed is considered as the output one. The value of  $ A = max_j A_j$ can be regarded as the alternative definition of classifier fitness.

Let us notice that in order to calculate the fitness, we have to study the evolution of the network on all elements of the training dataset. It means that the training dataset should not be too large and, moreover, that we need a fast algorithm to describe the network time evolution, including possible interaction between oscillators. 

When we decide about the definition of the classifier fitness, we can initiate the evolutionary optimization \cite{foge94,weic07}  of the initially generated population of $M$ classifiers. The next generation of a population of classifiers is obtained as the results of the following operations  \cite{gizy16}:
The upper $5\%$ of the most fit classifiers were copied to the next generation. 
The remaining  $95\%$ of elements of the next generation were generated by recombination and mutation processes applied to pairs of classifiers randomly selected from the upper  $50\%$  of the fittest ones.  The operations are illustrated in Fig. 3. Recombination of two Parents produces a single Offspring by combining random parts of their body \cite{gold89}.  Next, we applied mutation operations on the Offspring. We assumed that an input oscillator can change into a normal oscillator and vice versa. Also, illumination times of normal oscillators and the parameters of the function that translates the predictor value into the illumination time of an input oscillator $t_{start}, t_{end}$ are subject of mutations. The probabilities of mutations are selected such that, on average, a single mutation occurs in each clasifier within a single evolution step. Since the fitness changes during evolution, the position of the output is not fixed and also can change from generation to generation. 
We do not exclude the case in which an input droplet is used as the output one. As a result of trial and error, the fitness of the best classifier increases with the number of generations. The optimization was repeated until the preassumed final number of generations was reached.

The definitions of classifier fitness given above are not equivalent. The maximum fitness: $H(R)$ when the mutual information is used and $A = 1$ if we use the accuracy both describe a perfect classifier. However, the optimizations for   $I(R,O) $ and the optimization for $A$ are not equivalent. It can be shown that an increase in $I(R,O) $ can decrease $A$ and vice versa  \cite{gore20a}.

\begin{figure}
\centerline{\includegraphics[width=10cm]{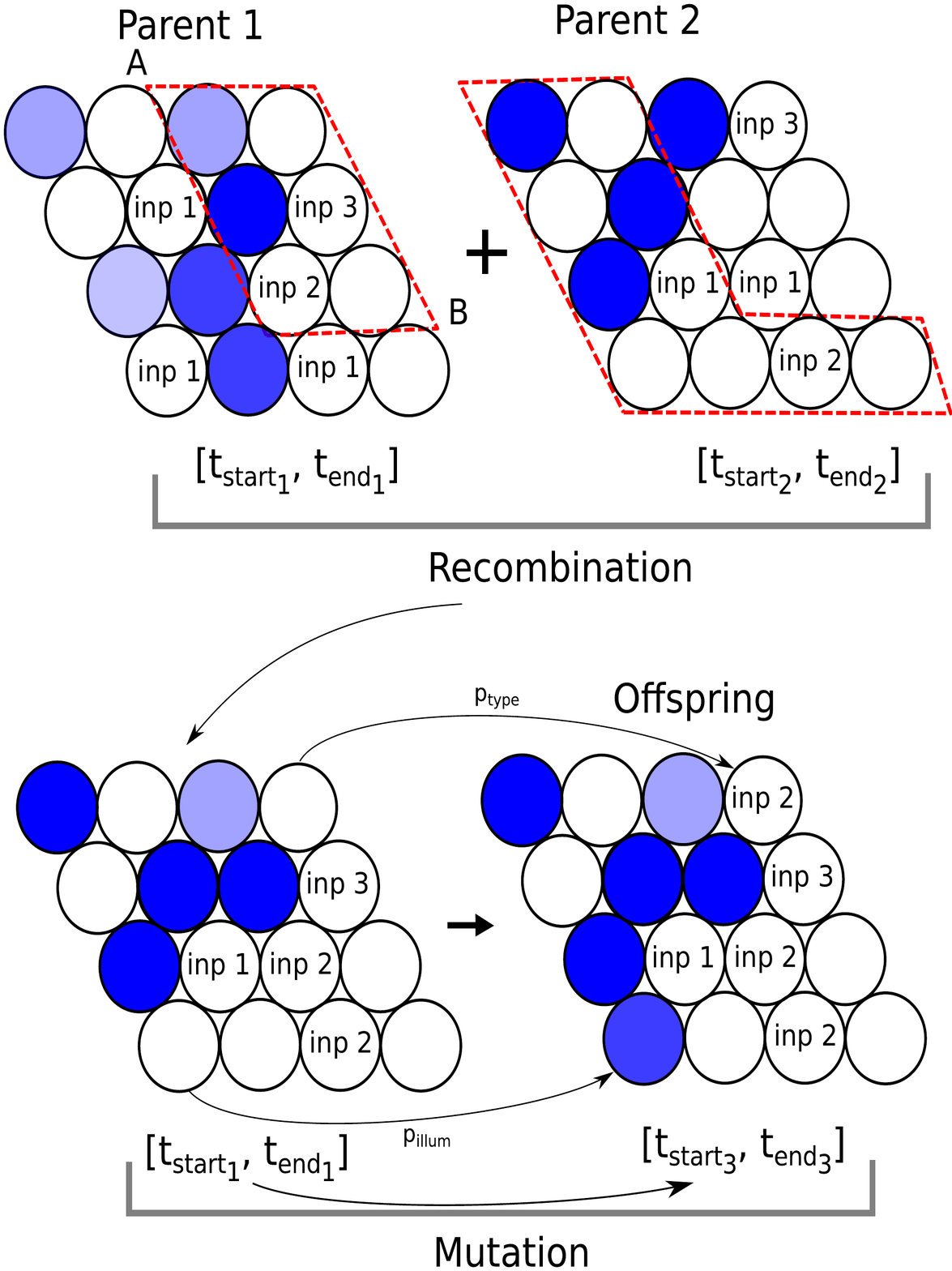} }
\caption{
A schematic illustration of steps leading to the generation of an Offspring from two Parents. First, the  Recombination step is performed:  
randomly selected points A and B in the structure of Parent 1 mark a parallelogram, which is copied, along with the illumination interval for inputs, to the Offspring. The other part of the Offspring comes from Parent 2. Then, during the Mutation of an Offspring, droplet types, and initial illumination times and the function that translates the predictor value into input droplet illumination time are subject to mutations. The intensity of blue color in each droplet indicates the illumination time for normal droplets. 
}
\label{ra_fig3}
\end{figure}

\clearpage

\section{Results}

Here we would like to answer questions that have not been addressed in the previous papers. In our first report on the SIC problem \cite{gizy16} we estimated the network accuracy on the basis of results obtained for the training dataset. Test datasets of different sizes, different from the training dataset, were used in \cite{zomm19} to estimate the classifier accuracy. As can be expected, the accuracy estimated using the training dataset was higher than the values obtained with the test datasets. Here, in order to obtain more precise information on the size of the dataset needed for accurate estimation of classifier quality, we calculated the mutual information and the accuracy using a few testing datasets with the same number of elements. The comparison between the mutual information and the accuracy obtained for the training dataset and for testing datasets of different sizes is presented in  Fig. 4.  We considered a network of 7 oscillators arranged in a hexagonal geometry  shown in Fig. 4A. The blue numbers identify individual oscillators. The input oscillators are marked by ''in x'',   ''in y''  and  ''in z''.  The same identification of oscillators is used in Figs. 5,6, and 7. The touching oscillators are interacting. The illumination of the central, normal oscillator was $t_{illum}(3) = 11.95 s$. The values of parameters describing the relationship between  were: $t_{start} =   68.44 s$ and $  t_{end} = 83.63$. The structure shown in Fig. 4A  was obtained after $700$ evolutionary steps of optimization towards the maximum accuracy. The population of $M=1000$ classifiers was optimized using a training dataset of $K=1000$ elements.    The optimization program produced a network structure with an interesting symmetry.  The inputs of $x$ (the oscillator \#6), of $y$ (the oscillator \#1), and of $z$  (the oscillator \#2) are symmetrically distributed around the central oscillator that is the output one. These inputs are separated by the inputs of the $z-$ coordinate (oscillators \#0,4 and 5).  In Fig. 4A, the blue color intensity on normal oscillators increases with their illumination time. The mutual information $ I(R,O_j)$ is marked with red color in the form of a pie
chart where the sector size is normalized to the maximal value of mutual information - the entropy of training dataset ($0.992 $ bit). The
output droplet is marked with a wide black border. Figure 4B illustrates the distribution of excitation numbers observed on the central oscillator (\#3) for all records from the training dataset. The red and green bars correspond to points located inside and outside the sphere, respectively. As seen in Fig 4B for the majority of points located inside the sphere the network produced 3 excitations on the central oscillator. On the other hand, for the majority of points located outside the sphere 4 excitations of the central oscillator were observed. The mutual information between the outputs of the training dataset and the number of excitation is $0.397$ (the large black dot in Fig. 4D).  The classification rule: if 3 excitations at the central oscillator are observed, then the point is located within the sphere and if 4 excitations at the central oscillator are observed, then the point is located outside the sphere leads to the accuracy of $0.849$  (the large black dot in Fig. 4E).  Figure 4C illustrates the distribution of excitation numbers observed on the oscillator \#1 for all records from the training dataset. Again if 3 excitations were observed, then the probability of a record corresponding to a point inside the sphere was larger than the probability of a record corresponding to a point outside the sphere. On the other hand, if 4 excitations were observed, then with a high probability, we can claim that the processed record corresponded to a point outside the sphere. However, the accuracy of the classifier with the output oscillator \#1 is low and equals $0.656$. In order to see if the mutual information and the accuracy are correctly estimated using the training dataset containing only $1000$ records, we have calculated both quantities using testing datasets of different sizes in the range from $2000$ to $50 000$ records. We used $3$ different test datasets for each size, and the results are marked by dots of different colors in Figs. 4D, E. Moreover, we calculated the mutual information and the accuracy using the test dataset of $500,000$ records. These results are considered as the most objective estimation of the classifier quality.  They are shown using a horizontal black line in Figs. 4 D, E.   The values obtained for the test dataset of $500,000$ records are slightly lower than those estimated for the training dataset. The results demonstrate that in the case of the SIC problem, in order get less than  $ 2\%$  error for the mutual information and less than $ 1\%$  error for the accuracy, one should use the test dataset of $50,000$ records or more.

\begin{figure}
\centerline{\includegraphics[width=10cm]{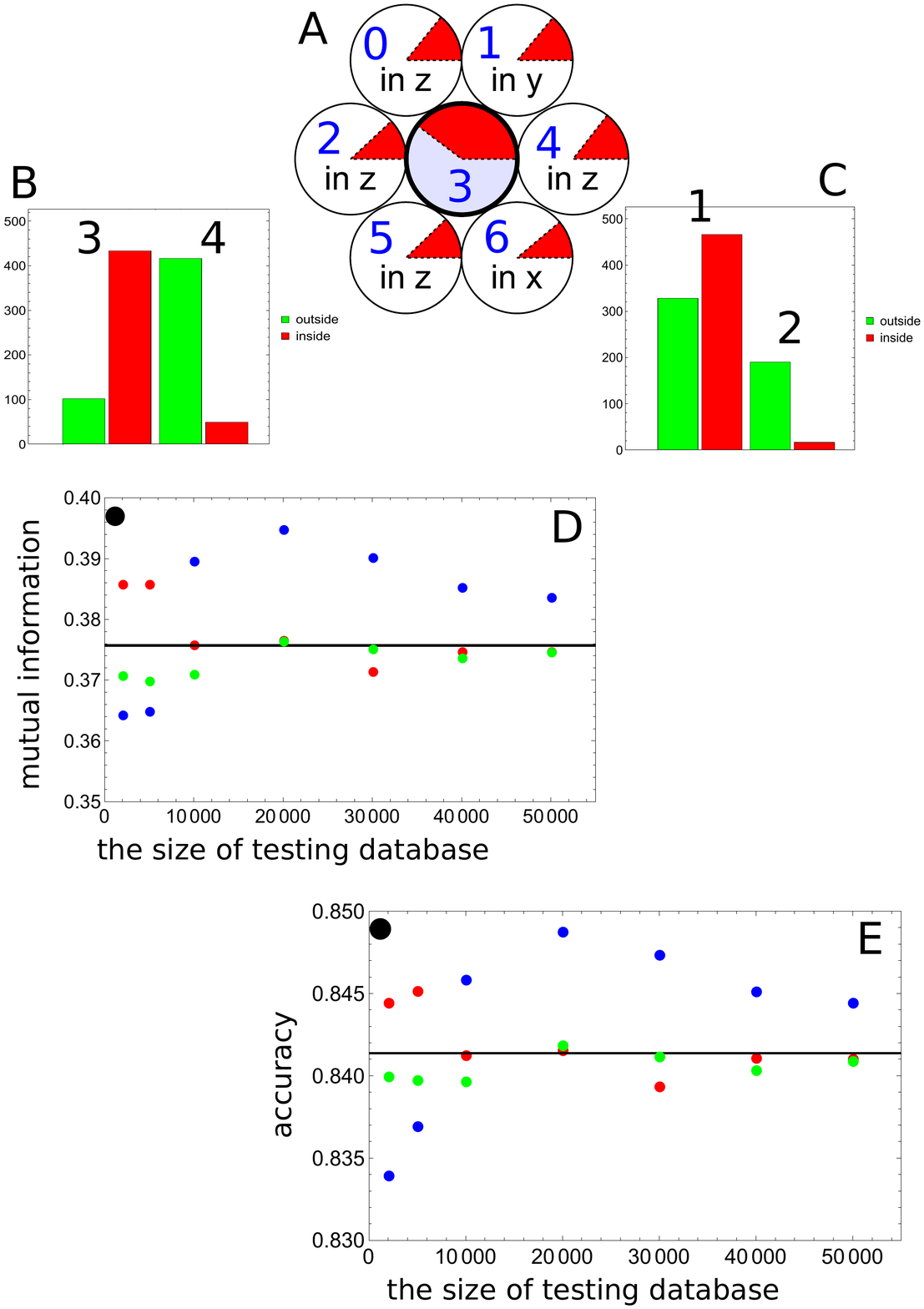} }
\caption{The comparison between the mutual information and the accuracy obtained for the training dataset and the testing datasets of different sizes. Fig.4 A shows the structure of the considered network. The blue numbers identify individual oscillators.  The touching oscillators are interacting. The blue color intensity corresponds to illumination time of the normal droplet as a fraction of $t_{max}$.  The mutual information $ I(R,O_j)$ is marked with red color in the form of a pie chart where the sector size is normalized to the entropy of the training dataset. The
output droplet is marked with a wide black border.
 Figures 4B and 4C  illustrate the distribution of excitation numbers observed on the oscillators \#3 and \#1  for all records from the training dataset. The red and green bars correspond to points located inside and outside the sphere, respectively. Figures 4D and 4E show the mutual information and the accuracy calculated for the training dataset (large black dots), 3 different testing datasets of different sizes in the range from $2000$ to $50 000$ records (red, blue, and green dots), and a large test dataset of $500 000$ records (the black line). 
}
\label{ra_fig4}
\end{figure}

Now let us consider the problem of the size of the training dataset needed for the optimization of a classifier of the SIC problem. In evolutionary optimization, we have to calculate the fitness of each classifier in the population. In order to calculate the fitness, we have to study the classifier answer to all elements of the training dataset. Therefore, assuming a fixed size of the classifier population, the numerical complexity of optimization linearly depends on the number of records in the training dataset. In calculations, we considered the classifier composed of $7$ oscillators in the geometry illustrated in Figure 4 A. We considered training datasets with different numbers of records $K$ from $100$ to $2000$. In each case, the population of $M=1000$ classifiers was optimized for $40000$ evolutionary steps to maximize the accuracy. In both Figs. 5A and 5B  the line color codes the size of the training dataset $K$.
 In all cases, the major increase in both accuracy and mutual information is observed within the first $1000$ of evolution steps. Only a small improvement of the accuracy was observed at the later stage of optimization. The structures of optimized classifiers are shown in between Figs. 5A, and 5B.  The normal output oscillator is always in the center.  For all considered training datasets except of  $K=500$  we obtained structures symmetric with respect to $x- $, $y- $ and $z- $ inputs. The other parameters describing the presented classifiers are:\\
- for  $K=100$:  $t_{start} =   67.32 s$, $  t_{end} = 86.07$, $t_{illum}(0) = 51.92 s$, $t_{illum}(3) = 38.15 s$, $t_{illum}(4) = 38.62 s$, $t_{illum}(5) = 35.23 s$. The accuracy measured using the  training dataset was $0.890$ and $I(R,O) = 0.513$.\\
- for  $K=200$:  $t_{start} =    67.18 s$, $  t_{end} =  84.81$, $t_{illum}(0) =  19.51 s$, $t_{illum}(3) = 17.86 s$, $t_{illum}(4) = 43.35 s$, $t_{illum}(5) = 25.40 s$. The accuracy measured using the  training dataset was $0.900$ and $I(R,O) = 0.533$.\\
- for  $K=500$:  $t_{start} =   67.68 s$, $  t_{end} = 84.14$, $t_{illum}(0) =  51.92 s$, $t_{illum}(3) = 11.94 s$. The accuracy measured using the  training dataset was $0.0.882$ and  $I(R,O) = 0.482$.\\
- for  $K=1000$:  $t_{start} =   67.76 s$, $  t_{end} = 83.07$, $t_{illum}(3) =11.94 s$. The accuracy measured using the  training dataset was $0.881$ and  $I(R,O) = 0.478$.\\
- for  $K=2000$:  $t_{start} =   68.11 s$, $  t_{end} = 85.09$, $t_{illum}(0) = 52.94 s$, $t_{illum}(3) = 21.43 s$, $t_{illum}(4) = 29.33 s$, $t_{illum}(5) = 14.31 s$. The accuracy measured using the  training dataset was $0.870$ and  $I(R,O) = 0.445$.\\

The fact that the accuracy was larger for small training datasets can be explained by the fact that it is easier to get a high accuracy for a small training dataset.
If the accuracy was evaluated using a large testing dataset ($50000$ records; as  seen in Fig. 4 E the evaluation with a test dataset of this size was expected to be objective) we obtained: $ 0.812, 0.863, 0.857, 0.874$ and $ 0.860$ for $K = 100, 200, 500, 1000$ and $K=2000$ respectively.  Therefore, as expected, the accuracy of the classifier optimized with the smallest training dataset was the smallest one. On the other hand, the qualities of all other classifiers are at a similar level. We conclude that the training dataset of $ K \ge 200$ elements includes a sufficient number of records to reflect the SIC problem correctly. 

\begin{figure}
\centerline{\includegraphics[width=10cm]{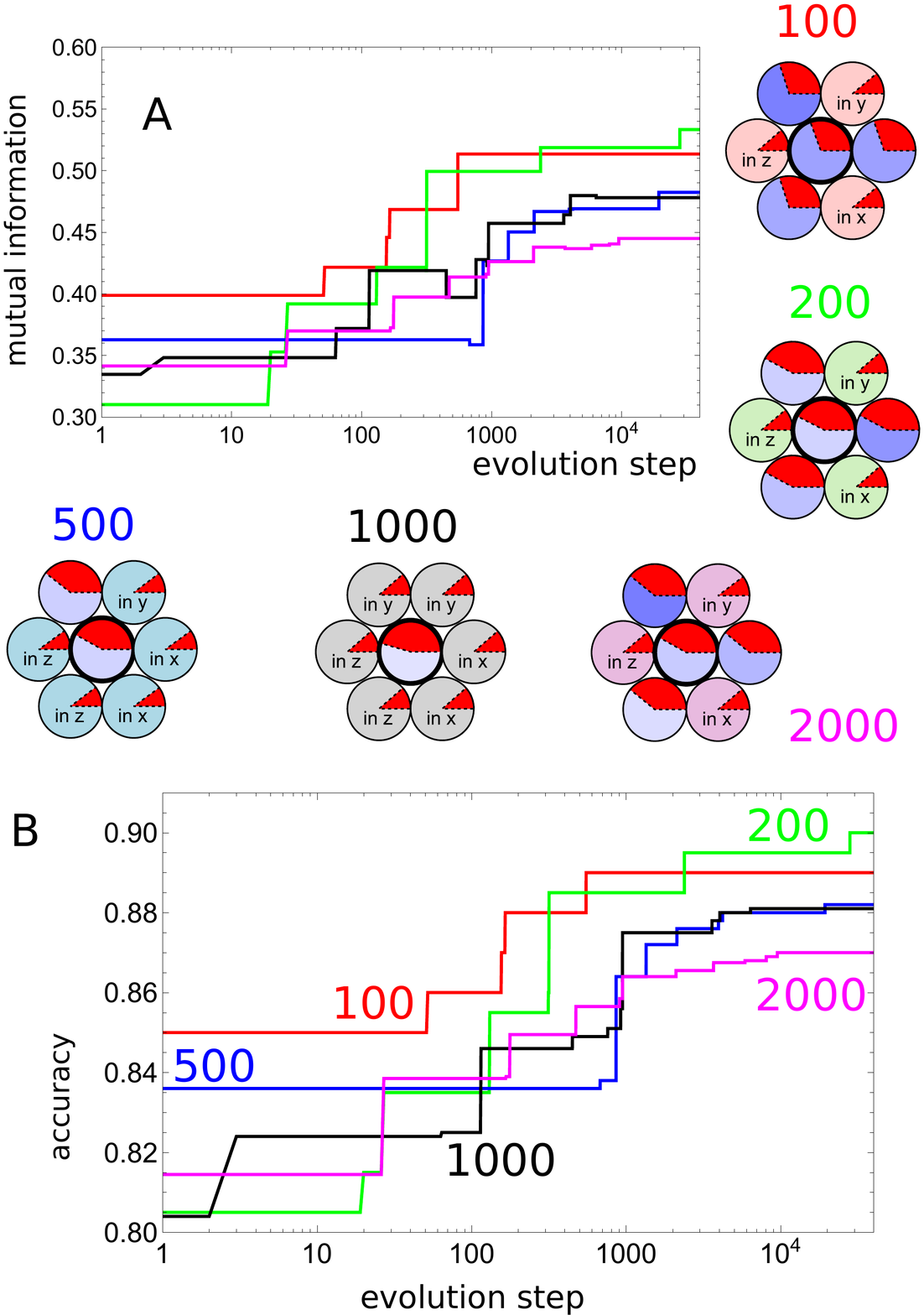} }
\caption{The progress of the classifier optimization for different sizes of the training dataset from $K=100$ to $K=2000$ records.  The population of $M=1000$ individuals was used. The accuracy was optimized for $40000$ of evolution steps. Figs. 5 A and 5 B show the mutual information and the accuracy as functions of the number of evolutionary steps. In both figures, the line color codes the size of the training dataset. The structures of classifiers obtained for different values of $K$ are shown in between Figs. 5A, and 5B. }
\label{ra_fig5}
\end{figure}

The numerical complexity of the optimization program linearly depends on the size of the optimized population $M$.  However, it can be expected that optimization is more efficient if the population is large. We investigated classifier optimization for $M$ in the range from $50$ to $500$   considered populations of $50, 80, 100, 200$ and $500$ classifiers.
Figure 6  illustrates the mutual information (A) and the classifier accuracy (B) as functions of the number of evolution steps for different numbers of elements included in the considered population. In calculations, we considered the classifier composed of $7$ oscillators in the geometry illustrated in Figure 4 A. In each case, the classifiers were optimized for $20000$ evolutionary steps to maximize accuracy. The training dataset of $K=1000$ records was used.  In both Figs. 6A and 6B  the line color codes the size of optimized population $M$. Like in Fig. 5, the rapid increase in the accuracy is observed at the beginning of optimization, here within the first $2000$ steps of evolution. The parameters describing the optimized classifiers are:\\
- for  $M=50$:  $t_{start} =  23.40  s$, $  t_{end} = 66.52$, $t_{illum}(1) = 39.45 s$, $t_{illum}(2) =  52.82 s$, $t_{illum}(3) = 5.62 s$, $t_{illum}(6) = 12.88  s$. The accuracy measured using the  training dataset was $0.813$ and  $I(R,O) = 0.331$ .\\
- for  $M=80$:  $t_{start} =  41.39 s$, $  t_{end} =  58.63 $, $t_{illum}(1) =  49.66 s$, $t_{illum}(3) =  27.71 s$, $t_{illum}(5) = 46.91 s$, $t_{illum}(6) = 53.19 s$. The accuracy measured using the  training dataset was $ 0.875$ and  $I(R,O) =  0.462$ .\\
- for  $M=100$:  $t_{start} =   67.14 s$, $  t_{end} = 83.67$, $t_{illum}(2) =  44.33 s$, $t_{illum}(3) = 19.30 s$, $t_{illum}(4) =12.46 s$, $t_{illum}(6) = 11.52 s$.  The accuracy measured using the  training dataset was $0.0.881$ and  $I(R,O) = 0.474$ .\\
- for  $M=200$:  $t_{start} =   67.67 s$, $  t_{end} = 85.78$, $t_{illum}(1) =8.57 s$, $t_{illum}(2) =14.07 s$, $t_{illum}(3) =20.18 s$, $t_{illum}(6) =16.01 s$. The accuracy measured using the  training dataset was $0.869$ and $I(R,O) = 0.439$ .\\
- for  $M=500$:  $t_{start} =  67.63 s$, $  t_{end} = 85.46$, $t_{illum}(0) =13.65 s$, $t_{illum}(3) = 22.84 s$, $t_{illum}(4) = 19.29 s$, $t_{illum}(5) = 29.69 s$. The accuracy measured using the  training dataset was $0.874$ and  $I(R,O) = 0.453$ .\\

 It can be seen that both the mutual information ( $ 0.33$)  and the classifier accuracy ($0.813$)  
obtained for the population of $M=50$ classifiers are significantly lower than those obtained for the other populations. For the majority of populations, the optimized structures are symmetrical. Non-symmetrical structures were observed for populations of $80$ and $100$ classifiers. Nevertheless, both the accuracy and the mutual information for $M=80$ and $100$ measured on the training datasets are similar to those observed for classifiers optimized with populations of $M= 200, 500$ and $M= 1000$ shown in Fig. 5.  

 The values of  the mutual information  and the classifier accuracy decrease if we use a large testing dataset (here of $500 000$ elements). For this test dataset we obtain: $A= 0.785$ and  $I(R,O) =  0.284$ for $M=50$, 	 
$A=  0.857$ and  $I(R,O) =  0.410$ for $M=80$, 
$A=   0.864$ and  $I(R,O) =  0.434$for $M=100$, 
$A=  0.864$ and     $I(R,O) = 0.425$ 	 for $M=200$, 	 
$A= 0.868$ and      $I(R,O) = 0.437$  for $M=500$.	 
These results suggest that within the used optimization procedure, the optimized population should contain $M \ge 100$ elements to achieve the goal.

\begin{figure}
\centerline{\includegraphics[width=10cm]{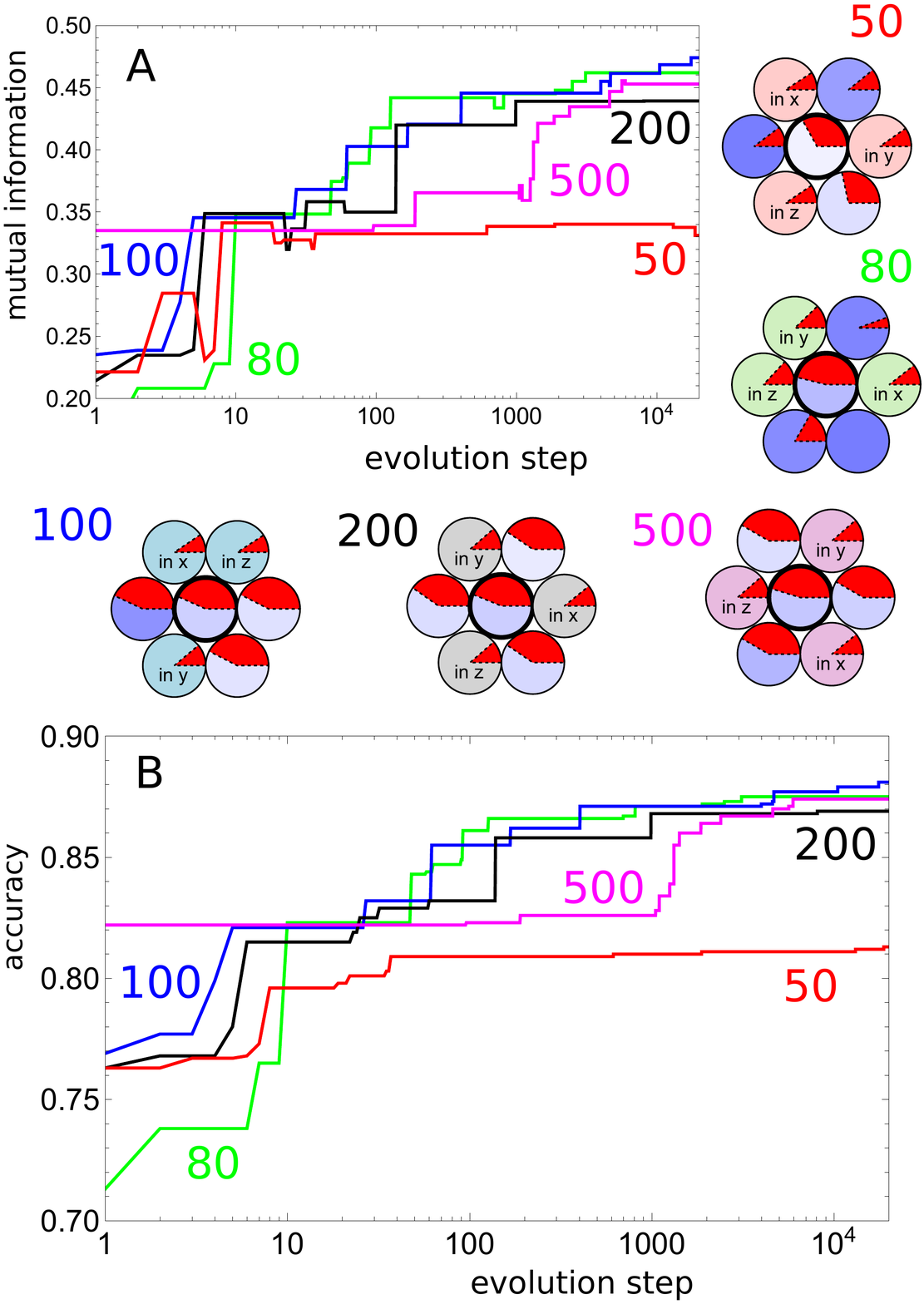} }
\caption{The progress of the classifier optimization for different populations of classifiers  from $M = 50$ to $M = 500$.  The training dataset of $K=1000$ records was used. The accuracy was optimized for $20000$ of evolution steps. Figs. 6 A and 6 B show the mutual information and the accuracy as functions of the number of evolutionary steps. In both figures, the line color codes the size of the training dataset. The structures of classifiers obtained for different values of $K$ are shown in between Figs. 6A, and 6B. 
}
\label{ra_fig6}
\end{figure}

Finally, we studied the problem if the accuracy of a classifier for the SIC problem can be improved by selecting a special geometry of the network.
Figure 7 presents results for networks of classifiers characterized by different geometries. We optimized parameters of a small (T4) and a large triangle (T10), of a David shield structure made of 13 oscillators (DS), and of a regular geometry of 4x4 = 16 oscillators (REG). 
 The considered geometries are shown in the mid of the figure. The numbers inside the circles identify individual oscillators. We assume that touching oscillators interact. Almost all considered geometries  (excluding the regular one) allow for a symmetrical distribution of the inputs to the anticipated output oscillator located in the center. For T4, the lines indicate interactions between the oscillators located at the triangle corners.  The line color in Fig. 7A and Fig 7B codes the network type.  In each case, the population of $M=1000$ classifiers was optimized using the training dataset of $K=1000$ records for $20000$ evolutionary steps to maximize the mutual information.
 The parameters describing the optimized classifiers are:\\
- for T4:  $t_{start} =  67.75  s$, $  t_{end} = 83.63$, $t_{illum}(1) = 11.73 s$. The classifiecation rule is: 3 excitations on the output oscillator represent a point inside the sphere, 4 excitations correspond to a point outside the sphere.  The accuracy measured using the  training dataset was $0.884$ and  $I(R,O) = 0.487$.\\
- for T10:  $t_{start} =  67.85 s$, $  t_{end} =  82.86 $, $t_{illum}(4) = 41.32 s$, $t_{illum}(6) =  31.72 s$, $t_{illum}(8) =  23.43 s$. The classifiecation rule is: 2 excitations on the output oscillator represent a point inside the sphere, 3 excitations correspond to a point outside the sphere.  The accuracy measured using the  training dataset was $ 0.889$ and  $I(R,O) =  0.497$.\\
- for DS:  $t_{start} =   82.68 s$, $  t_{end} =  67.93 $, $t_{illum}(4) =  31.13 s$, $t_{illum}(6) =  11.44 s$, $t_{illum}(9) = 46.60 s$, $t_{illum}(11) =  45.23 s$.  The classifiecation rule is: 3 excitations on the output oscillator represent a point inside the sphere, 4 or 5 excitations correspond to a point outside the sphere.   The accuracy measured using the  training dataset was $0.891$ and  $I(R,O) = 0.504$.\\
- for REG:  $t_{start} =   83.34 s$, $  t_{end} =67.10$, $t_{illum}(0) =41.38 s$, $t_{illum}(1) =54.76 s$, $t_{illum}(3) =8.80 s$, $t_{illum}(4) =26.31 s$, $t_{illum}(9) =42.07 s$, $t_{illum}(11) =57.47 s$, $t_{illum}(12) =28.50 s$, $t_{illum}(13) =11.62 s$, $t_{illum}(14) =16.08 s$, $t_{illum}(15) =48.18 s$.  The classifiecation rule is: 4 excitations on the output oscillator represent a point inside the sphere, 5 or 6 excitations correspond to a point outside the sphere.The accuracy measured using the  training dataset was $0.903$ and $I(R,O) = 0.554$.\\

 For almost all considered geometries (except REG), the mutual information was close to its maximum value within the first 1000 steps of optimization, as it was observed in the results presented in Figs. 5 and 6. The values of mutual information and of the accuracy, measured on the training dataset, show the anticipated increase with the number of oscillators forming the classifier. The trend is confirmed by the values of  $I(R,O)$ and the accuracy $A$ obtained using a large test dataset ($500000$ records). Here  $I(R,O)=  0.453,  0.468,   0.470$ and $0.513$ for T4, T10, DS and REG respectively. Similarly, the values of $A$ are $0.872,  0.877,  0.880$ and $ 0.893$. Therefore we can see that both the mutual information and the accuracy are increasing with the number of oscillators forming the network. Nevertheless, the increase is slow if compared with the increase in the number of oscillators considered and related increase in the numerical complexity of optimization.

\begin{figure}
\centerline{\includegraphics[width=10cm]{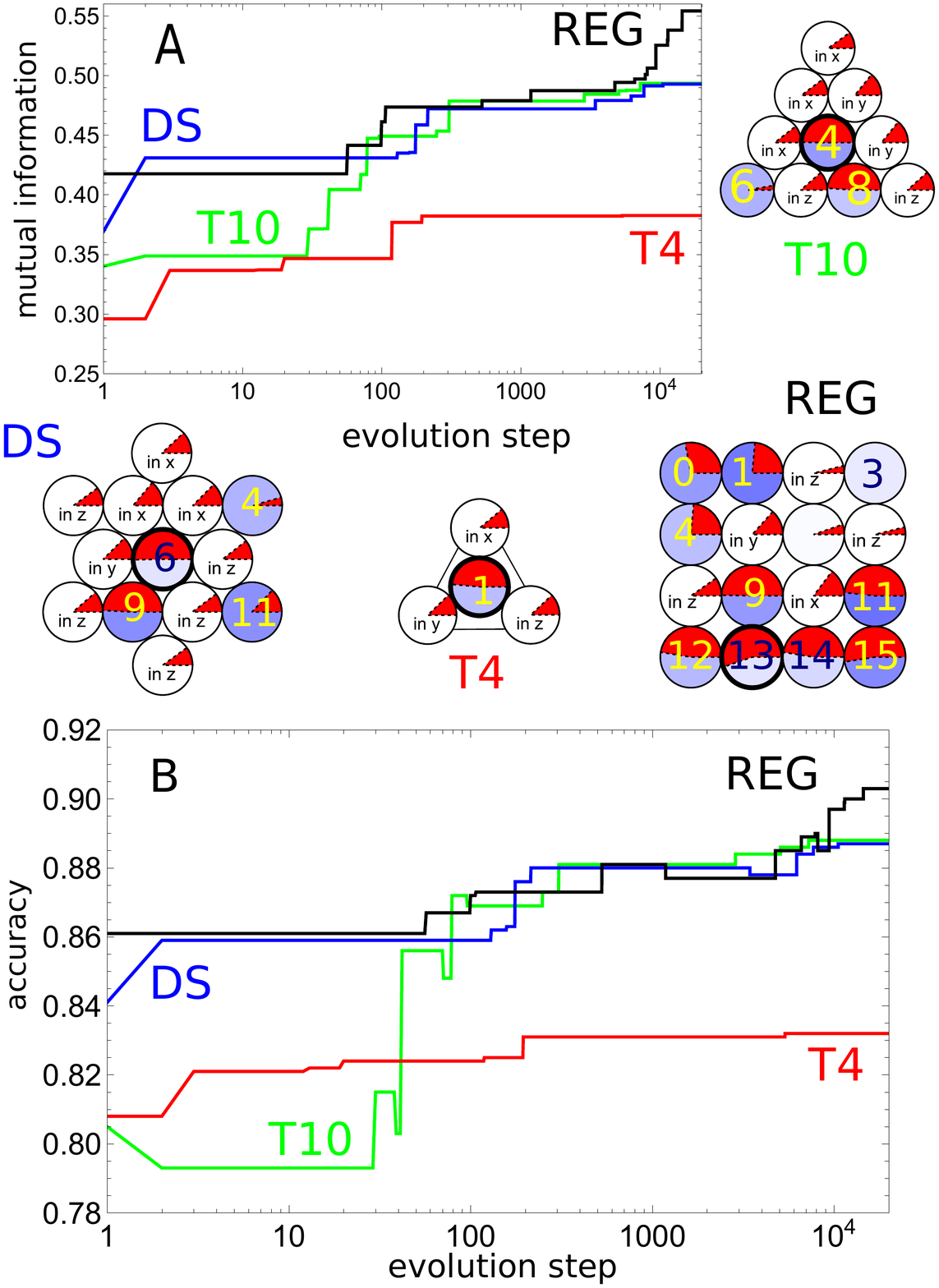} }
\caption{
The progres of the classifier optimization for classifiers characterized by different network geometry.
The population of $M=1000$ classifiers was considered, and the training dataset of $K=1000$ records was used. The population was optimized for the maximum mutual information for $20000$ of evolution steps. Figs. 7 A and 7 B show the mutual information and the accuracy as functions of the number of evolutionary steps.  In both figures, the line color codes the network geometry. The structures of classifiers obtained for different geometries are shown in between Figs. 7A, and 7B. The numbers inside the circles identify individual oscillators.
}
\label{ra_fig7}
\end{figure}

\clearpage

\section{Conclusions}

In the Chapter, we have answered a few technical questions related to the optimization of classifiers for the Sphere In a Cube problem that are based on a network of chemical oscillators. In our simulations, we used a simplified and fast to simulate event-based-model illustrated in Fig. 1. Network optimization was done by a simple evolutionary algorithm. 
We have found that it is relatively easy to obtain a classifier that determines the point location with an accuracy higher than 85 \%, and such accuracy can be achieved by a network formed by only 4 oscillators.
The number of records in the training dataset $K$ and the size of the optimized population $M$  are important parameters of the program and they have a direct impact on the execution time. We have found that reasonable results for the SIC problem can be obtained when the both  $K$ and $M$ are in the order of several hundred. However, for objective estimation of classifier accuracy (say with an error below 1\%) the classifier should be tested on much larger datasets containing $K \ge 50000$ records.

It is worthwhile to mention that we did not observe a significant  influence of the size of a dataset used for optimization on the classifier accuracy. The results obtained for training datasets in the range between 200 and 2000 records were almost identical. This can be explained using the following argument. It seems that networks of the considered medium  are not able to classify correctly the Sphere In a Cube problem because nor of the classifiers showed accuracy exceeding 90\%. To achieve the accuracy of 85+ \% it is not necessary to use a training database that describes the problem with high accuracy. For such accuracy, an approximation of Sphere In a Cube problem by a training dataset of 200 records seems to be sufficient to introduce the correlations in the oscillator network that reflect the data structure. 

The improvement of classifier accuracy by selecting the proper network geometry was the main motivation for our study. We  considered networks of different geometry, and some of them  reflected the symmetry of the problem.  The results are rather negative. Nor of the structures  seem to perform much better than the others. In all cases, we achieved accuracy close to $\sim 87 \%$, but we were not able to increase the accuracy above $90 \%$. We believe that the limitation can be linked to unflexibility of the event-based-model. The future improvment of chemical classifiers seem to be possible if we ue more realistic models of oscilaltors and intractions between them, that include parametrs allowing controll on these processes.

\clearpage

\section{Acknowledgements}
The work was supported by the Polish National Science Centre grant UMO-2014/15/B/ST4/04954.


\end{document}